\documentclass[lettersize,journal]{IEEEtran}
\usepackage{amsmath,amsfonts}
\usepackage{algorithmic}
\usepackage{algorithm}
\usepackage{array}
\usepackage{makecell}
\pdfoutput=1
\usepackage[caption=false,font=small,labelfont=rm,textfont=rm]{subfig}
\usepackage{textcomp}
\usepackage{stfloats}
\usepackage{url}
\usepackage{verbatim}
\usepackage{graphicx}
\usepackage{cite}
\usepackage{caption}
\newcommand{\secref}[1]{Section~\ref{#1}}
\newcommand{\figref}[1]{Fig.\ref{#1}}
\newcommand{\tabref}[1]{Table~\ref{#1}}
\newcommand{\equref}[1]{(\ref{#1})}
\begin{document}

\title{Bias-Compensated State of Charge and State of Health Joint Estimation for Lithium Iron Phosphate Batteries}

\author{Baozhao Yi,~Xinhao Du,~Jiawei Zhang,~Xiaogang Wu,~\IEEEmembership{Member,~IEEE,}~Qiuhao Hu,~Weiran Jiang,~Xiaosong Hu,~\IEEEmembership{Senior Member,~IEEE,}~and~Ziyou Song,~\IEEEmembership{Member,~IEEE}

\thanks{This work was supported by the Singapore MOE Tier 1 project under Grant A-8001187-00-00 (Corresponding authors: Xiaogang Wu and Ziyou Song).}
\thanks{B. Yi, J. Zhang and Z. Song are with Department of Mechanical Engineering, National University of Singapore, 117575, Singapore (emails: e1010669@u.nus.edu, jiawei.zhang@u.nus.edu and ziyou@nus.edu.sg).}
\thanks{X. Du is with School of Electrical and Electronic Engineering, Harbin University of Science and Technology, Harbin, 150080, China (email: hpudxh@163.com).}
\thanks{X. Wu is with School of Electrical Engineering, Hebei University of Technology, Tianjin, 300130, China (email: 2023051@hebut.edu.cn).}
\thanks{Q. Hu and W. Jiang are with Farasis Energy USA, Inc., Hayward, 94545, CA, USA (emails: qhhu@umich.edu and wjiang@farasis.com).}
\thanks{X. Hu is with Department of Mechanical and Vehicle Engineering, Chongqing University, Chongqing, 400030, China (email: xiaosonghu@ieee.org).}}



\maketitle

\begin{abstract}
Accurate estimation of the state of charge (SOC) and state of health (SOH) is crucial for the safe and reliable operation of batteries. Voltage measurement bias highly affects state estimation accuracy, especially in Lithium Iron Phosphate (LFP) batteries, which are susceptible due to their flat open-circuit voltage (OCV) curves. This work introduces a bias-compensated algorithm to reliably estimate the SOC and SOH of LFP batteries under the influence of voltage measurement bias. Specifically, SOC and SOH are estimated using the Dual Extended Kalman Filter (DEKF) in the high-slope SOC range, where voltage measurement bias effects are weak. Besides, the voltage measurement biases estimated in the low-slope SOC regions are compensated in the following joint estimation of SOC and SOH to enhance the state estimation accuracy further. Experimental results indicate that the proposed algorithm significantly outperforms the traditional method, which does not consider biases under different temperatures and aging conditions. Additionally, the bias-compensated algorithm can achieve low estimation errors of below 1.5\% for SOC and 2\% for SOH, even with a 30mV voltage measurement bias. Finally, even if the voltage measurement biases change in operation, the proposed algorithm can remain robust and keep the estimated errors of states around 2\%.
\end{abstract}


\section{Introduction}\label{section 1}
\IEEEPARstart{A}{s} the global energy crisis intensifies, electric vehicles (EVs) are emerging as a crucial solution \cite{blomgrenDevelopmentFutureLithium2017}. Lithium-ion batteries (LIBs) are the primary power sources for EVs, with their high energy density and long cycle lives. However, the heedless operation of LIBs without rigorous battery management can lead to significant performance degradation and increased safety risks. As a result, accurate battery state estimation within battery management systems (BMSs) is crucial for ensuring the safety of EVs. Two primary battery states are estimated by the BMS: State of Charge (SOC) and State of Health (SOH). Specifically, SOC represents the ratio between the remaining capacity of the battery and its total available capacity \cite{TRUCHOT2014218}, while SOH refers to the proportion of the battery's maximum available capacity to its nominal capacity \cite{liuResearchBatterySOH2022}. Besides the undeniable importance of SOC and SOH estimation, it is imperative to accurately estimate the parameters of the battery model. This confirms the reliability of model-based algorithms, especially in light of parameter variations across diverse operational conditions and aging phases \cite{xiongLithiumionBatteryAging2020}.

Various model-based battery models have been proposed, including the Equivalent Circuit Model (ECM) \cite{huComparativeStudyEquivalent2012} and the electrochemical model \cite{kleinElectrochemicalModelBased2013}. Moreover, several data-driven battery models have also been introduced, including the Convolutional Neural Network (CNN) model \cite{shenDeepLearningMethod2019}, the Long Short-Term Memory (LSTM) model \cite{liukailong9040661}, etc.

In addition to various battery models, many advanced estimation algorithms have also been proposed. The Coulomb Counting method \cite{ngEnhancedCoulombCounting2009} is an open-loop approach to SOC estimation. However, its performance is dependent on the accuracy of the initial SOC. With its burgeoning popularity, The Extended Kalman Filter (EKF) \cite{plettExtendedKalmanFiltering2004} stands as a prominent tool for SOC estimation. While EKF's reputation in SOC estimation is well-acknowledged, the Jacobian matrix computation poses challenges, especially when simultaneously estimating states and parameters. To address these problems, a novel sequential algorithm is designed in \cite{songSequentialAlgorithmCombined2020}. By employing high-pass filters and injecting signals of varying frequencies, battery parameters are estimated sequentially using EKFs. This approach refines computational processes and enhances the precision of estimating parameters and states. There are other estimation techniques, such as unscented Kalman filter \cite{heStateChargeEstimation2013}, adaptive filter \cite{hongwenheStateofchargeEstimationLithiumion2011}, etc. Given the intertwined nature of SOC and SOH estimation processes, a multi-scale EKF has been proposed in \cite{huMultiscaleFrameworkExtended2012}, \cite{xiongDatadrivenMultiscaleExtended2014}. Furthermore, a dual fractional-order EKF is proposed for the estimation of both SOC and SOH simultaneously, which not only improves the convergence speed but also improves the precision of the estimation \cite{huCoestimationStateCharge2018}.

While research into both battery models and estimation algorithms is essential, the significance of data quality must be addressed. Three critical benchmarks for data evaluation emerge---the sensitivity analysis \cite{laiAnalyticalSensitivityAnalysis2019}, the Fisher information matrix \cite{sharmaFisherIdentifiabilityAnalysis2014} and the Cramér-Rao (CR) bound \cite{SCHARF1993301}. A data selection strategy in \cite{linDataSelectionStrategy2018} aims to enhance the precision of battery parameter estimation by exclusively selecting high-sensitivity data. The CR bound, derived from the inversion of the Fisher information matrix \cite{linAnalyticAccuracyBattery2016}, sets the stage in \cite{songParameterIdentificationLithiumion2018} for the designs of optimal current profiles. Besides, academic investigations have validated that the accuracy achieved in estimating a single parameter often exceeds that achieved when estimating all states and parameters simultaneously based on the theory of CR bound \cite{songCombinedStateParameter2020}.

Although current estimation algorithms, such as the EKF, and data evaluation techniques, such as the CR bound, have helped mitigate estimation errors of battery states, they still have some limitations, primarily based on their assumptions. These methods assume that voltage measurement errors are solely attributable to variance and tend to overlook the potential impact of measurement bias. For instance, the CR bound represents the smallest variance of estimation error achievable when measurement bias is absent. Meanwhile, the EKF assumes that measurement error follows a Gaussian distribution centered around zero. In reality, voltage measurement bias, often caused by the sensor drift, can also compromise the accuracy of SOC estimation \cite{ZHENG2018161}, especially for Lithium Iron Phosphate (LFP) batteries, which display relatively flat OCV-SOC curves within the mid-SOC range. Since the EKF assumes its estimation results are unbiased, it suggests that estimated error includes only estimation variance, without considering estimation bias. Nevertheless, when there is a bias in the voltage measurement, it is important to understand whether the estimated SOC is biased. In general, the flat OCV-SOC curve observed in LFP cells within the mid-SOC range is often described as an ``error amplifier" \cite{zhengCellStateofchargeInconsistency2013}, as it makes state estimation highly sensitive to noise and bias.

To address the challenges associated with voltage measurement bias, this study introduces a bias-compensated algorithm that aims to ensure accurate estimation of SOC and SOH, focusing on mitigating the effects of voltage measurement bias on LFP batteries. The contents of this paper are as follows. In \secref{section 2}, the battery system dynamic is introduced. After that, the estimation processes of battery parameters can be separated by adding high-pass filters and injecting signals with different frequencies \cite{songCombinedStateParameter2020}, which prevents the parameter estimation process from being influenced by voltage measurement bias. In \secref{section 3}, first, we examine how voltage measurement bias affects the estimation process of SOC. Based on the analysis, the estimated error of SOC caused by voltage measurement bias is negligible if the slope of the OCV-SOC curve is high. According to this finding, the framework of the ``bias-compensated algorithm'' is proposed to improve the estimation accuracy of SOC and SOH in the presence of voltage measurement bias. The core of the proposed algorithm is estimating SOC and capacity using a Dual Extended Kalman Filter (DEKF) only in the high-slope SOC regions. Alternatively, for the SOC range with a relatively small slope, capacity is not updated and SOC is calculated by ampere-hour integration method. Besides, the voltage measurement biases are estimated in the low-slope SOC intervals, and their estimated values are then compensated in the next joint estimation of SOC and capacity to improve the state estimation accuracy further. Experimental results in \secref{section 4} prove the superiority of the proposed algorithm, compared to the traditional method, without considering different SOC intervals and bias compensation. Even if the voltage measurement bias is 30mV, the estimated errors from the bias-compensated algorithm are still low, with below 1.5\% for SOC and 2\% for capacity. Moreover, the proposed algorithm can maintain its stability, keeping the errors of states at approximately 2\% when the voltage biases change during operation. Finally, the conclusion can be found in \secref{section 5}.

\begin{figure}[t]
\centering
    \includegraphics[scale=0.8]{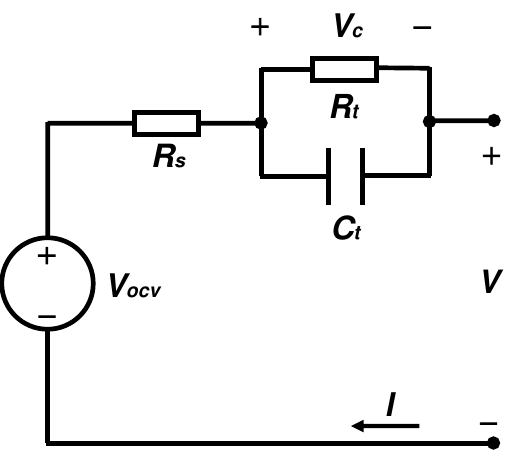}
    \caption{The first-order equivalent circuit model.}
    \label{fig1}
    \vspace{-10pt}
\end{figure}

\section{Battery system modeling}\label{section 2}
\subsection{Battery system description}\label{section 2.1}
The first-order ECM \cite{tranComparativeStudyEquivalent2021} is adopted in this study to strike a balance between estimation fidelity and computational efficiency. According to \figref{fig1}, the dynamic of the battery system is given as:
\begin{equation}
    \label{space}
    \left\{
    \begin{aligned}
        \dot{V}_c &= -\frac{1}{R_tC_t}V_c + \frac{1}{C_t}I\\
        V_b &=V_{ocv}-R_sI-V_c
    \end{aligned}
    \right.,
\end{equation}
where $R_s$, $R_t$, and $C_t$ represent the ohmic resistance, diffusion resistance, and capacitance, respectively. The voltage term generated by the RC pair is symbolized as $V_c$, and $V_b$ represents the battery terminal voltage. Furthermore, $V_{ocv}$ is the OCV. The relationship between $V_{ocv}$ and SOC can be represented by polynomial:
\begin{equation}
    \label{V_ocv}
    V_{ocv}  = A_{12}SOC^{12} + A_{11}SOC^{11} + \dots + A_1SOC +A_0.
\end{equation}
In this paper, we use $\rm 12^{th}$ polynomial to fit the OCV-SOC curve, and $A_{0-12}$ are polynomial coefficients. For small SOC changes, the relationship between $V_{ocv}$ and SOC can be linearized as follows \cite{dongKalmanFilterOnboard2016}:
\begin{equation}
    \label{simply}
    V_{ocv} = aSOC + b,
\end{equation}
where $a$ and $b$ are linearization coefficients to be fitted. The linearized function of the OCV-SOC curve will be used to simplify the analysis in the following sections, and \equref{V_ocv} will be applied for estimation. The SOC dynamic is given as follows:
\begin{equation}
    \label{SOC-capa}
    SOC  = SOC_0 - \int_{t_0}^{t}\frac{\eta}{Q_b}I(t)dt,
\end{equation}
where $SOC_0$, $\eta$, $Q_b$, and $t_0$ refer to initial SOC, charge/discharge efficiency, battery capacity, and initial time, respectively.

In addition to SOC and capacity, the model parameters should also be estimated online, including $R_s$, $R_t$, and time constant $\tau$, which refers to the product of $R_t$ and $C_t$. Two assumptions are made in the estimation process:
\begin{itemize}
    \item The initial $V_c$ is zero.
    \item $R_s$, $R_t$ and $\tau$ are constant in the short term (e.g., 20 mins).
\end{itemize}

If the voltage sensor includes a measurement bias, $\Delta V$, that is caused by extreme operation conditions like violent temperature changes, then the output voltage $V$ will be destroyed. The compromise voltage can be expressed as:
\begin{equation}
    \label{measured V}
    V = V_b + \Delta V.
\end{equation}
In this work, $\Delta V$ represents the mean of the voltage measurement error \cite{ZHENG2018161} and is assumed to be constant in a period  \cite{vemuriSensorBiasFault2001}.

Combining \equref{space} and \equref{simply}, the Laplace function of \equref{measured V} is given as:
\begin{equation}
    \label{space laplace}
    \begin{split}
        V(s) &= \left[\frac{aSOC_0}{s} + \frac{b}{s}\right] - \left[\frac{a}{s}\frac{\eta}{Q_b}I(s)\right] - \left[R_sI(s)\right]
        \\[2mm]
        &- \left[\frac{R_t}{1+\tau s}I(s)\right] + \left[\frac{\Delta V}{s}\right],
    \end{split}
\end{equation}
where $s$ is a complex Laplace variable. From \equref{space laplace}, it can be seen that there are five voltage terms, including the initial SOC term, SOC change term ($Q_b$ term), $R_s$ term, RC pair term, and $\Delta V$ term.
 
\subsection{The analysis of battery system dynamics}\label{section 2.2}
In \equref{space laplace}, both the initial SOC term and $\Delta V$ term remain constant. If a high-pass filter is applied to the battery system over time, these two voltage terms will gradually decay to zeros. The filtered system can be written as:
\begin{equation}
    \label{filtered}
    \begin{split}
        V_f(s) &= \frac{(aSOC_0 + b)T_c}{1 + T_cs} - \frac{a}{s}\frac{\eta}{Q_b}I_f(s) - R_sI_f(s)
    \\[2mm]
    &- \frac{R_t}{1 + \tau s}I_f(s) + \frac{\Delta VT_c}{1 + T_cs},
    \end{split}
\end{equation}
where $1/T_c$ is the cut-off frequency of the high-pass filter. $V_f$ and $I_f$ are given as:
\begin{align*}
        V_f(s) &= \frac{T_cs}{1 + T_cs}V(s), & I_f(s) &= \frac{T_cs}{1 + T_cs}I(s)
\end{align*}
where $V_f$ and $I_f$ are voltage and current after filtering, respectively. When changing \equref{filtered} into the time domain, the initial SOC term and $\Delta V$ term will become as follows:
\begin{equation}
    \label{decay}
    \left\{
    \begin{aligned}
        &L^{-1}\left[\frac{(aSOC_0 + b)T_c}{1 + T_cs}\right] = (aSOC_0 + b)e^{-\frac{t}{T_c}} \to 0\\
        &L^{-1}\left[\frac{(\Delta V)T_c}{1 + T_cs}\right] = \Delta Ve^{-\frac{t}{T_c}} \to 0
    \end{aligned},
    \right.
\end{equation}
which will be removed from the battery system over time. Consequently, only $R_s$, RC pair, and $Q_b$ terms exist in the filtered system and need to be estimated. In our previous study \cite{songCombinedStateParameter2020}, these three parameters can be estimated sequentially by introducing high-frequency and medium-frequency signals sequentially, and applying the high-pass filters can shield the effects of voltage measurement bias on the estimation processes based on the analysis above.

\section{The Bias-compensated algorithm for joint estimation of SOC and SOH}\label{section 3}
\subsection{The effect of voltage measurement bias on SOC estimation}\label{section 3.1}
While high-pass filters can eliminate the effect of voltage measurement bias on parameter estimation, they are inaccessible to be used in SOC and capacity estimation due to the low-frequency nature of the initial SOC term and $Q_b$ term. Due to the coupled estimation processes of SOC and $Q_b$, the latter process mainly relies on the voltage changes resulting from SOC changes. Therefore, in this section, we only analyze how $\Delta V$ affects the SOC estimation accuracy by assuming the capacity is known.

The estimation results from the EKF are typically considered unbiased, assuming a zero-mean measurement noise. Now, we will explore whether the estimation result of SOC can remain unbiased in the presence of a constant measurement bias in the voltage data. The discrete state-space equation at this time can be expressed as:
\begin{equation}
    \label{discrete state-space}
    \left\{
    \begin{aligned}
        &SOC(k) = SOC(k-1) - \frac{\eta T_s}{Q_b}I(k) + w(k-1)\\
        &V(k) = a[SOC(k)] + b - R_sI(k) -V_c(k) + \Delta V + v(k)
    \end{aligned},
    \right.
\end{equation}
where $w$ and $v$ are process noise and measurement noise. $k$ is time step and $T_s$ refers to sampling period. $V_c(k)$ can be calculated as:
\begin{equation}
\label{V_c calculation}
    V_c(k) = e^{-\frac{T_s}{\tau(k)}}V_c(k-1) + R_t(k)\left(1 - e^{-\frac{T_s}{\tau(k)}}\right)I(k).
\end{equation}

Since the estimated values of $R_t$, $\tau$ are known and the initial $V_c$ is assumed to be zero, the value of $V_c(k)$ can be confirmed. The estimated outcome of SOC at $k^{th}$ time step is calculated as:
\begin{equation}
\label{SOC calculation}
    \hat{SOC}(k) = \hat{SOC}^{-}(k) + K(k)\left[V(k) - \hat{V}^-(k)\right],
\end{equation}
where $K(k)$ is Kalman gain at $k^{th}$ time step. The subscript \textasciicircum and $-$ represent the estimation and prior value, respectively. $\hat{SOC}^{-}(k)$ is given as:
\begin{equation}
\label{SOC- calculation}
    \hat{SOC}^{-}(k) = \hat{SOC}(k-1) - \frac{\eta T_s}{Q_b}I(k).
\end{equation}
$\hat{V}^-(k)$ of \equref{SOC calculation} is expressed as:
\begin{equation}
\label{V(k)-}
    \hat{V}^-(k) = a[\hat{SOC}^{-}(k)] + b - R_sI(k) -V_c(k).
\end{equation}

By taking \equref{SOC- calculation} into \equref{SOC calculation} and subtracting  \equref{SOC calculation} from the state function of \equref{discrete state-space}, we can get the estimated error $e(k)$ of $SOC(k)$:
\begin{equation}
\label{e(k)}
\begin{split}
    e(k) &= \left[1 - K(k)a\right]e(k-1) + \left[1 - K(k)a\right]w(k-1) 
    \\[2mm]
    &- K(k)\left[\Delta V + v(k)\right].
\end{split}
\end{equation}

Next, we will calculate the expectation of \equref{e(k)}, $E\left[e(k)\right]$, to check whether the estimation error of SOC remains unbiased. If $E\left[e(k)\right]$ equals zero, the estimation error is unbiased. Conversely, any other value indicates that the estimation error becomes biased. As the means of $w(k-1)$ and $v(k)$ are assumed to be zero \cite{kalmanfilter}, and $\Delta V$ is a constant, $E\left[e(k)\right]$ can be transformed as follows:
\begin{equation}
\label{E[e(k)]}
    E\left[e(k)\right] = \left[1 - K(k)a\right]E\left[e(k-1)\right] - K(k)\Delta V.
\end{equation}

When the initial estimated error of SOC is $e(0)$, $E\left[e(k)\right]$ can be changed into:
\begin{equation}
\label{E[e(k)]_c}
\begin{split}
    E\left[e(k)\right] = E_1 - E_2,
\end{split}
\end{equation}
where $E_1$ and $E_2$ refers to:
\begin{equation}
    \label{E_1 and E_2}
    \left\{
    \begin{aligned}
        &E_1 = \left\{\prod_{i=1}^k\left[1-K(i)a\right]\right\}E\left[e(0)\right]\\[2mm]
        &E_2 = \left\{\sum_{i=1}^k\left[\prod_{j=i+1}^k(1-K(j)a)K(i)\right]\right\}\Delta V \nonumber
    \end{aligned}.
    \right.
\end{equation}

There are two non-zero terms in $E\left[e(k)\right]$, in which the first term $E_1$ is related to the expectation of $e(0)$ and the second component $E_2$ is decided by $\Delta V$. As $k$ tends towards infinity, $E_1$ will converge to zero because $\left[1-K(i)a\right]$ is consistently less than 1. Moreover, the Kalman gain $K$ converges to a constant because of the linear relationship between SOC and OCV \cite{linAnalyticDerivationBattery2017}. When time is infinite, the \equref{E[e(k)]_c} can be expressed as follows:
\begin{equation}
\label{E[e(k)]_f}
    E\left[e(k)\right]\ \to\ \frac{\Delta V}{a}.
\end{equation}

In conclusion, we have proved that the estimation error of SOC is biased since $E\left[e(k)\right]$ is non-zero. Our analysis findings align with those presented in  \cite{linTheoreticalAnalysisBattery2018}. Based on the information in \equref{E[e(k)]_f}, the estimated SOC obtained by the EKF cannot converge to the true SOC since there is an error term related to the slope of OCV-SOC curve and the voltage measurement bias. When the slope of the OCV-SOC curve is steep, $E\left[e\left(k\right)\right]$ tends to be small, resulting in more accurate SOC estimation. However, for LFP batteries, there is a flat zone in the OCV-SOC curve from 10\% to 95\% SOC with a small slope. Therefore, in this particular scenario, it is not advisable to estimate SOC and $Q_b$ since the value of $E\left[e(k)\right]$ is large, indicating a significant error in SOC estimation.

\begin{figure*}[t]
\centering
    \includegraphics[scale=0.35]{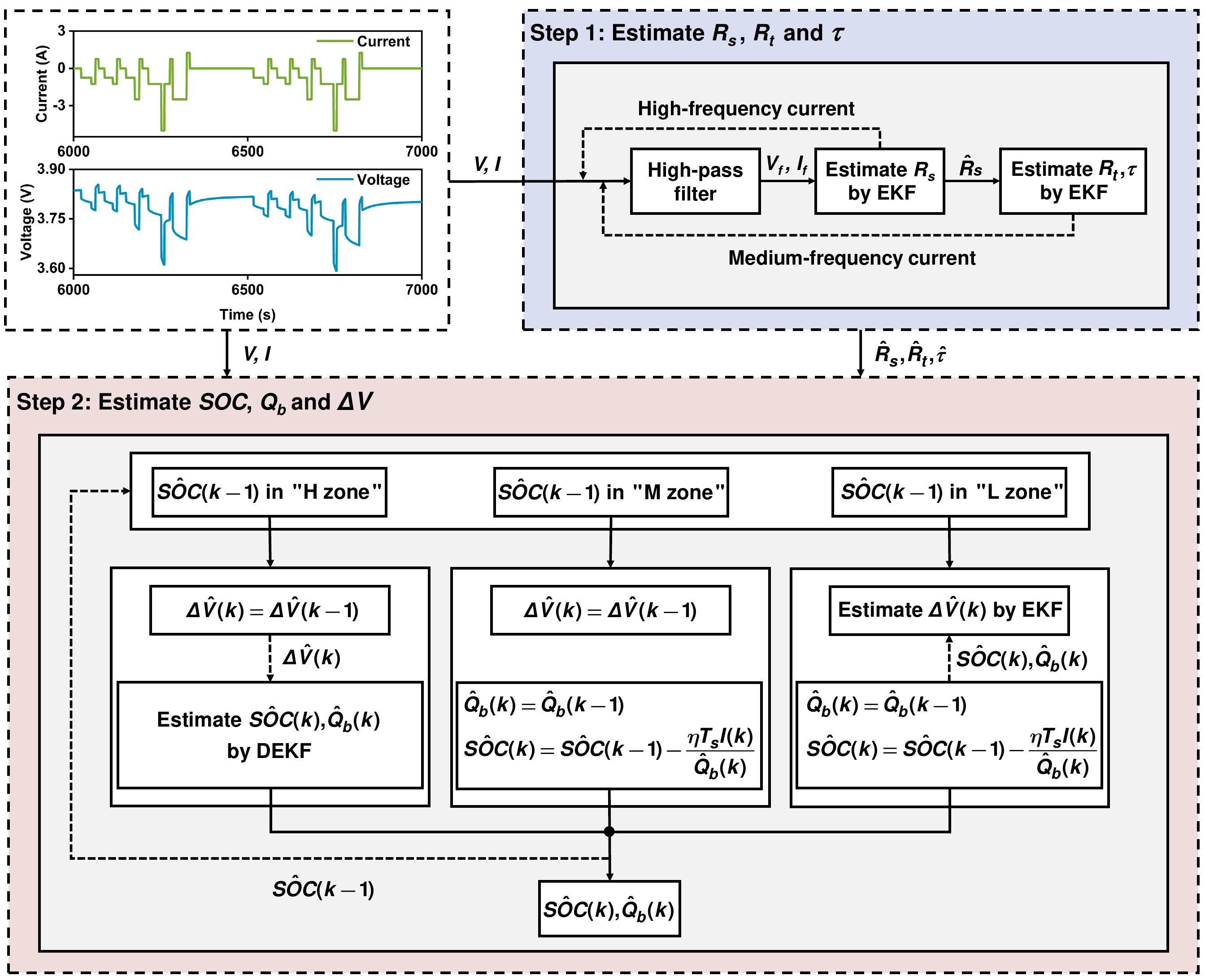}
    \caption{The framework of the bias-compensated algorithm.}
    \label{fig2}
    \vspace{-10pt}
\end{figure*}

\subsection{The framework of the bias-compensated algorithm}\label{section 3.2}
In \secref{section 3.1}, we have proved that a measurement bias in voltage data inevitably introduces an estimation bias in SOC estimation, impairing its accuracy. To mitigate the adverse effects of voltage measurement bias on the estimation of battery parameters and states, we propose a bias-compensated algorithm. This approach combines the high-pass filters and the injections of signals with different frequencies to estimate $R_s$, $R_t$, and $\tau$ through the EKFs. Subsequently, these estimated results will be used as the foundation for further estimating of SOC, $Q_b$, and $\Delta V$.

Moreover, when SOC lies within the ``High-slope zone" (or ``H zone")—a region characterized by a steep slope in the OCV-SOC curve, the DEKF is employed to estimate both SOC and $Q_b$. A comprehensive description of the DEKF's estimation process can be found in our prior work \cite{songSequentialAlgorithmCombined2020}. The ``H zone" is not arbitrarily defined; it is intended to confine the estimation of SOC and $Q_b$ in the high-slope region of SOC. This approach is supported by the insights from \secref{section 3.1}, revealing that the estimated errors of SOC caused by $\Delta V$ are smaller if the slope of OCV-SOC curve is large. Typically, for most LFP batteries, the SOC range of 0-10\% can be categorized as the ``H zone". Besides, within this zone, the slope of OCV-SOC increases as the SOC declines. When the SOC equals 0, the slope is greater than 15V. $\Delta V$ is not estimated at this time.

Conversely, when the SOC is in the ``Low-slope zone" (or ``L zone"), characterized by a gentle slope in the OCV-SOC curve below 0.05V, the EKF is used to estimate the voltage measurement bias, $\Delta V$. This estimated value is then compensated in the subsequent combined estimation process for SOC and $Q_b$. The formulation of the ``L zone" is primarily geared towards the precise estimation of $\Delta V$. The rationale is grounded in the minimal discrepancy in OCV, attributed to SOC estimation errors, thereby increasing the precision of the deduced $\Delta V$ value. Predominantly, for LFP batteries, the SOC range of 40\% to 50\% qualifies the ``L zone''. Within this interval, SOC is calculated via the ampere-hour integration method, and $Q_b$ remains unestimated.

SOC intervals are not from the ``H zone" or ``L zone" are considered as the ``Medium-slope zone" (or ``M zone"), distinguished by an intermediate OCV-SOC curve slope. Within the ``M zone", SOC is estimated through the ampere-hour integration method, and both $Q_b$ and $\Delta V$ remain unchanged. The detailed process of the bias-compensated algorithm is described as follows.

\textbf{Step 1:} The first step is estimating $R_s$, $R_t$ and $\tau$. This estimation process can be divided into two distinct stages. The first stage introduces a high-pass filter and injecting the high-frequency signal to estimate $R_s$. By doing so, the terminal voltage is dominated only by $R_s$ dynamic \cite{songSequentialAlgorithmCombined2020}. The battery terminal voltage can be simplified as:
\begin{equation}
    \label{estimate R_s}
    V_f(s) = -R_sI_f(s).
\end{equation}

With this simplification, we can now write the discrete-time state-space function as follows:
\begin{equation}
    \label{state-space_Rs}
    \left\{
    \begin{aligned}
        &R_s(k) = R_s(k-1) + r(k-1)\\
        &V_f(k) = -R_s(k)I_f(k) + v(k)
    \end{aligned},
    \right.
\end{equation}
where $r(k)$ is the process noise for parameters. $v(k)$ is the voltage measurement noise. The second stage of this step is adding another high-pass filter and injecting a medium-frequency signal to estimate $R_t$ and $\tau$. Currently, The voltage is governed by dynamics of $R_s$ and RC pair terms. Therefore, the \equref{filtered} can be simplified as:
\begin{equation}
    \label{estimate R_t}
    V_f(s) = -R_sI_f(s) - \frac{R_t}{1+\tau s}I_f(s).
\end{equation}

The estimated $R_s$ is used in this process, and the bilinear transformation is applied to discretize \equref{estimate R_t}. At this time, the state-space function is changed to:
\begin{equation}
    \label{state-space_Rt}
    \left\{
    \begin{aligned}
        &\boldsymbol{\theta}_2(k) = \boldsymbol{\theta}_2(k-1)+ \boldsymbol{r}(k-1)\\
        &V_f(k) = -R_s(k)I_f(k) -R_t(k)I_2(k) + v(k)
    \end{aligned},
    \right.
\end{equation}
where
\begin{equation}
    \label{I2}
    \left\{
    \begin{aligned}
        &\boldsymbol{\theta}_2(k) = \left[R_t(k)\ \tau(k)\right]^T\\
        &I_2(k) = \frac{T_s}{T_s + 2\tau}\left[I_f(k) + I_f(k-1)\right]-\frac{T_s - 2\tau}{T_s + 2\tau}I_2(k-1)\nonumber
    \end{aligned},
    \right.
\end{equation}
and $T_s$ is set to 1s. The estimated values of $R_s$, $R_t$, and $\tau$ will be used in the following step.

\textbf{Step 2:} This step can be divided into multiple parts, each corresponding to three SOC intervals with different slopes of the OCV-SOC curve. First, if $S\hat{O}C(k-1)$ is in the ``H zone'', SOC and $Q_b$ at $k^{th}$ time step are estimated by the DEKF. The state-space function is given as follows:
\begin{equation}
    \label{state-space_SOC}
    \left\{
    \begin{aligned}
        &Q_b(k) = Q_b(k-1) + r(k-1)\\
        &SOC(k) = SOC(k-1) - \frac{\eta T_s}{Q_b(k)}I(k) + w(k-1)\\
        &V(k) = OCV\left[SOC(k)\right] - R_s(k)I(k) - V_c(k) + \Delta V(k)
    \end{aligned},
    \right.
\end{equation}
where $OCV\left[\cdot\right]$ is the OCV-SOC function in \equref{V_ocv}. $V_c(k)$ is calculated by \equref{V_c calculation} and $\Delta V(k)$ is not updated during this time. $w(k-1)$ is the process noise for battery states. The initial value of $\Delta V$ is assumed to be unknown and set to 0. 

When $S\hat{O}C(k-1)$ is in ``M zone'' or ``L zone'', $SOC(k)$ will be calculated by the ampere-hour integration method that is written as follows:
\begin{equation}
    \label{Coulomb Counting}
    SOC(k) = SOC(k-1) - \frac{\eta T_s}{Q_b(k)}I(k).
\end{equation}
During this process, $Q_b(k)$ is not updated.

In addition, if $S\hat{O}C(k-1)$ is in ``L zone'', $\Delta V(k)$ is estimated by the EKF. The state-space function is given as follows.
\begin{equation}
    \label{state-space_SOC_bias}
    \left\{
    \begin{aligned}
        &\Delta V(k) = \Delta V(k-1) + w(k-1)\\
        &V(k) = OCV\left[SOC(k)\right] - R_s(k)I(k) - V_c(k) + \Delta V(k)
    \end{aligned}.
    \right.
\end{equation}

With time goes on, when $S\hat{O}C(k-1)$ is in ``H zone'' again, the estimated $\Delta V$ is taken into the \equref{state-space_SOC} to update the state-space function for estimating SOC and $Q_b$. The compensation of $\Delta V$ can enhance the accuracy of the state-space function used for estimating SOC and $Q_b$. This, in turn, leads to a more precise estimation of SOC and $Q_b$. Moreover, the improved accuracy in estimating SOC contributes to a more precise estimation of $\Delta V$. Finally, the accuracy of estimation for SOC, $Q_b$, and $\Delta V$ continues to improve as the algorithm iterates.

It should be noted that the bias-compensated algorithm necessitates the flexible adjustment of SOC zones according to the specific battery chemistry. In spite of that, the bias-compensated algorithm can be considered an approach for accurate state and parameter estimation of batteries with the influence of voltage measurement bias. The framework of the bias-compensated algorithm is shown in \figref{fig2}.

\section{Experimental results}\label{section 4}
\subsection{Experimental setup}\label{section 4.1}
To evaluate the practical performance of the proposed algorithm, a series of experiments are conducted. Two LFP batteries, which differ in terms of aging conditions, are used for experimental verification. Experiments are carried out under standard temperatures ($\rm 25^\circ C$) and lower temperatures ($\rm 5^\circ C$). The whole experimental verification is divided into four small experiments, which include Cell 1 at $\rm 25^\circ C$, Cell 1 at $\rm 5^\circ C$, Cell 2 at $\rm 25^\circ C$, and Cell 2 at $\rm 5^\circ C$. All details about the parameters of two batteries under different temperatures are presented in \tabref{table1}. The capacities of two LFP batteries at different temperatures are obtained from the capacity tests, and the `` true values'' of $R_s$, $R_t$ and $\tau$ are the results from Hybrid Pulse Power Characterization (HPPC) tests. The experimental bench includes an ARBIN BT2000 tester that generates current and voltage and a temperature chamber that provides constant temperature environments for batteries. Each experiment includes an initial stage (the initial charging stage) followed by four discharging/charging cycles. Throughout the process, the SOC ranges from 1\% to 70\%, including the ``H zone'', ``L zone'' and ``M zone''. A 0.1C constant current is applied to charge and discharge the battery model. After obtaining the voltage data, we intentionally introduce a 10mV measurement bias into the voltage data to represent the voltage bias in practice.

\begin{figure}[htbp]
\centering
\vspace{-10pt}
    \includegraphics[scale=1]{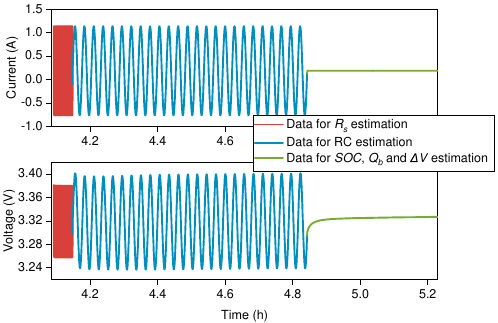}
    \caption{A piece of experimental data used for the estimation process.}
    \label{fig3}
\end{figure}

Within all experiments, the ``H zone" associated with the bias-compensated algorithm is from 0 to 10\% SOC, aligning with the peak of the OCV-SOC curve's slope. In contrast, the SOC range of 40\% to 50\% can be considered the ``L zone", reflecting the SOC region with the gentle slope. Other SOC intervals are ascribed to the ``M zone". The estimation process is as follows:

\begin{enumerate}
    \item First, the main task in the initial stage is estimating $R_s$, $R_t$, and $\tau$. As given in \cite{songSequentialAlgorithmCombined2020}, 0.5C signals of high (0.5Hz) and medium (0.01Hz) frequencies and high-pass filters are applied to estimate $R_s$, $R_t$ and $\tau$ at around 14400s in the initial stage. A piece of data used for estimating these three parameters in one experiment is shown in \figref{fig3}. Initial conjectures of $\hat{R}_s(0)$, $\hat{R}_t(0)$ and $\hat{\tau}(0)$ are 80\% of their own ``true values''. In addition, there is also a joint estimation of SOC and $Q_b$ before estimating these three parameters since $S\hat{O}C(k-1)$ is in the ``H zone'' at the beginning of the stage. The values of $S\hat{O}C(0)$ and $\hat{Q}_b(0)$ are posited as $\rm \left[10\%\ 8\ 0\right]^T$.
    \item The estimated $R_s$, $R_t$, and $\tau$ will be used in the following estimation processes of SOC, $Q_b$, and $\Delta V$. When $S\hat{O}C(k-1)$ is in ``H zone'', the DEKF is applied to estimate $SOC(k)$ and $Q_b(k)$. Conversely, if the $S\hat{O}C(k-1)$ occupies other zones, the $SOC(k)$ is calculated by the ampere-hour integration method while $Q_b(k)$ is not updated. Furthermore, as $S\hat{O}C(k-1)$ enters the ``L zone'' in the discharging stage of each cycle, the EKF will be used to estimate $\Delta V(k)$. And then, the estimated $\Delta V$ will be compensated to the subsequent joint estimation process of $SOC(k)$ and $Q_b(k)$.  In other SOC domains, barring the ``L zone'', $\Delta V(k)$ is not estimated. The value of $\hat{\Delta V}(0)$ is set to 0.
\end{enumerate}

To show the excellent performance of the bias-compensated algorithm, a comparison is conducted using the same current and voltage data. In this comparison, the SOC and $Q_b$ are continuously estimated by the DEKF without considering SOC ranges and bias compensation, i.e., termed continuous estimation in this paper. To better show the superiority of the bias-compensated algorithm, the true values of $R_s$, $R_t$, and $\tau$ are used for estimating SOC and $Q_b$ in the continuous estimation method.

To evaluate the estimation performance, the Root Mean Square Error (RMSE) of SOC and relative error (RE) of capacity are used, as calculated via \equref{RMSE} and \equref{RE}:
\begin{equation}
    \label{RMSE}
    RMSE = \sqrt{\frac{\sum_{k=1}^N\left[SOC_{t}(k)-S\hat{O}C(k)\right]^2}{N}},
\end{equation}
where the $SOC_{t}$ and $S\hat{O}C$ represent the true and estimated SOC, respectively. $N$ is the number of data points.
\begin{equation}
    \label{RE}
    RE = \frac{|\hat{Q}_b-Q_{b,t}|}{Q_{b,t}}\times100\%,
\end{equation}
where $\hat{Q}_b$ and $Q_{b,t}$ represent the estimated and true values of battery capacity, respectively.

\begin{table}[t]
    \centering
    \caption{Specification for two cells at different temperatures.}
    \begin{tabular}{|c||c||c||c||c|}
         \hline
         Conditions&$Q_b$ (Ah)&$R_s$ (m$\rm \Omega$)&$R_t$ (m$\rm \Omega$)&$\tau$ (s)\\
         \hline
         Cell 1 at $\rm 25^\circ C$&1.935&69&47&33\\
         \hline
         Cell 2 at $\rm 25^\circ C$&1.896&95&50&33\\
         \hline
         Cell 1 at $\rm 5^\circ C$&1.881&127&87&30\\
         \hline
         Cell 2 at $\rm 5^\circ C$&1.811&144&90&30\\
         \hline
    \end{tabular}
    \vspace{-15pt}
    \label{table1}
\end{table}

\subsection{Results and discussion}\label{section 4.2}
\begin{figure*}[htbp]
  \centering
  \includegraphics[scale=1]{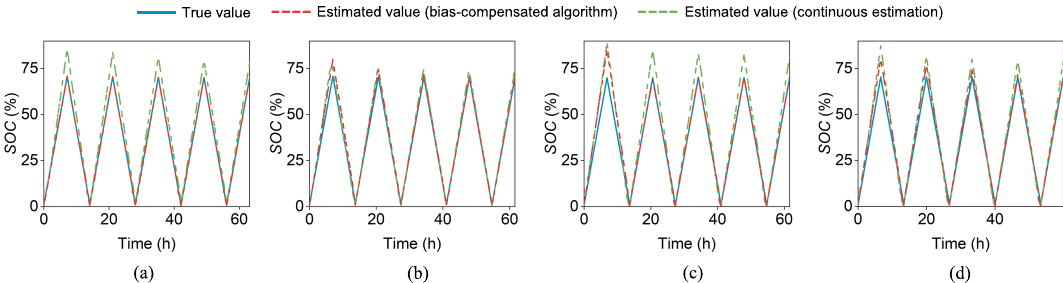}    
  \caption{Estimated results of SOC after adding voltage bias of 10mV. (a) Cell 1 at $\rm 25^\circ C$. (b) Cell 2 at $\rm 25^\circ C$. (c) Cell 1 at $\rm 5^\circ C$. (d) Cell 2 at $\rm 5^\circ C$.}
  \vspace{-10pt}
  \label{fig4}
\end{figure*}

\begin{figure*}[htbp]
  \centering
  \includegraphics[scale=1]{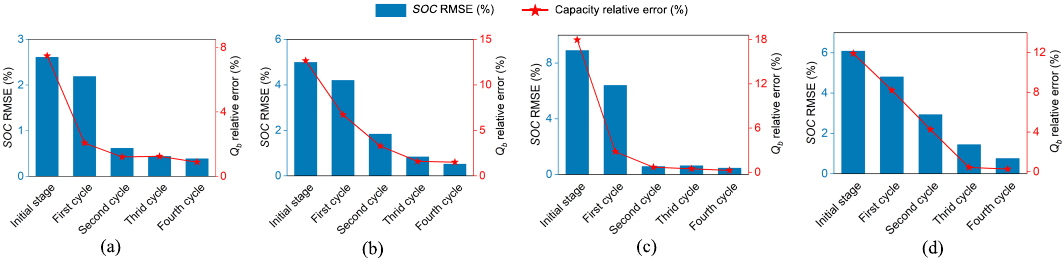}    
  \caption{Estimated errors of SOC and capacity in different stages after adding 10mV bias. (a) Cell 1 at $\rm 25^\circ C$. (b) Cell 2 at $\rm 25^\circ C$. (c) Cell 1 at $\rm 5^\circ C$. (d) Cell 2 at $\rm 5^\circ C$.}
  \vspace{-10pt}
  \label{fig5}
\end{figure*}
\figref{fig4} shows the estimation results of SOC after adding 10mV bias into voltage measurement. It can be observed that the continuous estimation method's output appears to struggle to estimate the true SOC accurately, particularly within certain SOC ranges where significant discrepancies are apparent. These observations show that the continuous estimation method's efficacy deteriorates significantly with the mere introduction of a 10mV bias, although the true values of $R_s$, $R_t$, and $\tau$ are used in its estimation process of SOC and capacity. \tabref{table2} presents estimated results of $R_s$, $R_t$, and $\tau$ obtained using the bias-compensated algorithm. It can be found that the estimated values of parameters differ from their ``true values''. Regarding the $R_s$, the "true values" are identified from HPPC tests and represent the values at 1Hz. Conversely, in our experiments, the estimated values for $R_s$ are estimated upon a sine current operating at the frequency of 0.5Hz. The Electrochemical Impedance Spectroscopy (EIS) in \cite{zhuNewLithiumionBattery2015} shows the differences of $R_s$ across different current frequencies. These differences become more pronounced as the temperature diminishes. Additionally, the estimated results of $R_t$ and $\tau$ also differ from their ``true values''. The primary error source is from $R_s$, which is necessary for estimating $R_t$ and $\tau$.

Nevertheless, the bias-compensated algorithm can overcome the model errors from inaccuracies of three parameters and achieve a more accurate estimation of SOC. Compared to the divergent results of continuous estimation, the estimated SOC from the bias-compensated algorithm has converged to the true SOC with negligible errors by the third discharging/charging cycle. \figref{fig5} provides more detailed information on the estimation errors of SOC and capacity in each discharging/charging cycle from the bias-compensated algorithm. It becomes evident that as the algorithm progresses, the estimation inaccuracies of SOC and $Q_b$ are gradually reduced. In the final stage of estimation in each experiment, the estimated error of SOC (i.e., SOC RMSE) is below 1\%. Furthermore, the estimated accuracy of $Q_b$ is also commendable, with all estimated errors of less than 2\%.

\begin{table}[b]
    \vspace{-15pt}
    \centering
    \caption{Estimation results of $R_s$, $R_t$ and $\tau$ from the bias-compensated algorithm.}
    \begin{tabular}{|c||c||c||c|}
         \hline
         Conditions&$R_s$ (m$\rm \Omega$)&$R_t$ (m$\rm \Omega$)&$\tau$ (s)\\
         \hline
         Cell 1 at $\rm 25^\circ C$&64.3&47&23.52\\
         \hline
         Cell 2 at $\rm 25^\circ C$&90.4&49.8&25.05\\
         \hline
         Cell 1 at $\rm 5^\circ C$&120.4&74.4&22.92\\
         \hline
         Cell 2 at $\rm 5^\circ C$&138.5&77.2&22.85\\
         \hline
    \end{tabular}
    \label{table2}
\end{table}

\begin{figure}[htbp]
  \centering
  \subfloat[]{\includegraphics[scale=1]{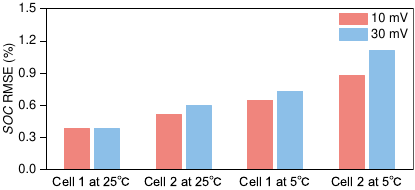}\label{fig6a}}
  \\
  \subfloat[]{\includegraphics[scale=1]{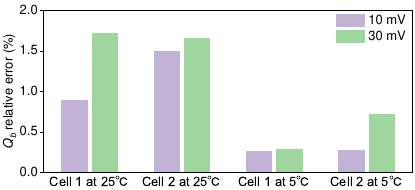}\label{fig6b}}
  \caption{SOC and capacity estimation errors. (a) RMSE of estimated SOC as compared to its true value. (b) RE between estimated $Q_b$ and its true value.}
  \vspace{-10pt}
  \label{fig6}
\end{figure}

Beyond the mere addition of a 10mV bias, we further challenge our algorithm by introducing biases of 30mV to assess its estimation prowess further. Please note that 30mV is far more than the common voltage bias in practice proposed in \cite{ZHENG2018161}. Given that the bias-compensated algorithm's estimation of SOC and $Q_b$ progressively track their true values over time, our analysis is refined to focus solely on the RMSE between the estimated and true SOC during the terminal discharging/charging cycle. \figref{fig6} shows the estimated errors of SOC and capacity when adding biases of 10mV and 30mV. As the bias in voltage measurements increases, the estimation inaccuracies for both SOC and $Q_b$ correspondingly amplify. However, in the face of these challenges, the bias-compensated algorithm consistently shows excellent estimation prowess. All RMSE of SOC are beneath 1.5\%, and the relative estimation errors for $Q_b$ are below 2\%.

\begin{figure}[t]
  \centering
  \includegraphics[scale=1]{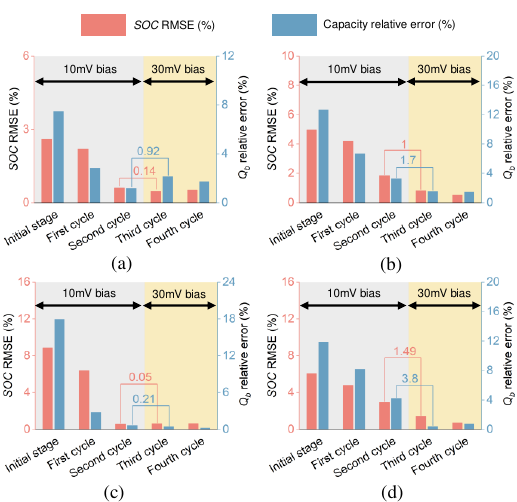}
  \caption{SOC and capacity estimation errors under the sudden bias change. (a) Cell 1 at $\rm 25^\circ C$. (b) Cell 2 at $\rm 25^\circ C$. (c) Cell 1 at $\rm 5^\circ C$. (d) Cell 2 at $\rm 5^\circ C$.}
  \vspace{-10pt}
  \label{fig7}
\end{figure}

In practice, the measurement bias is not always constant during operation. The bias of the voltage sensor may increase if it is suddenly exposed to an extreme operating condition. Next, we will check whether our proposed algorithm can maintain robustness when bias changes during operation. Assuming the voltage measurement bias is 10mV between the initial stage and the second discharging/charging cycle. And then, the bias will change to 30mV from the third to the fourth cycle. \figref{fig7} shows the estimated errors of the SOC and capacity through the whole process in each experiment. It is evident that the estimation accuracy of the SOC and capacity does not reduce due to the ``bias mutation''. Conversely, Three experiments, except `` Cell 1 at $\rm 25^\circ C$'' with slight error growths, still remain error decreases and the errors of SOC and capacity are around 2\% after the bias change. 

\begin{table}[b]
    \vspace{-15pt}
    \centering
    \caption{Estimated results of $\Delta V$ before and after the bias mutation.}
    \begin{tabular}{|c||c||c|}
         \hline
         \thead{Conditions}&\thead{Estimated values in \\ the second cycle (mV)}&\thead{Estimated values in \\ the third cycle (mV)}\\
         \hline
         Cell 1 at $\rm 25^\circ C$&0.154&20.225\\
         \hline
         Cell 2 at $\rm 25^\circ C$&-1.514&19.13\\
         \hline
         Cell 1 at $\rm 5^\circ C$&-2.745&16.875\\
         \hline
         Cell 2 at $\rm 5^\circ C$&-3.709&17.303\\
         \hline
    \end{tabular}
    \label{table3}
\end{table}

The robustness of the bias-compensated algorithm in this scenario benefits from the quick response to the bias variation. \tabref{table3} gives the estimation values of $\Delta V$ before and after the bias change. It can be found that the estimated biases have constant errors compared to their true values in each experiment. There are a few possible reasons to express this discovery. The first one is that the voltage sensor in the ARBIN BT2000 tester has inherent measurement biases. In addition, the estimated errors of biases are also caused by model errors, including inaccuracy of ECM, fitting and measured errors of OCV curves, and the estimated errors of $R_s$, $R_t$, and $\tau$. Since the absolute values of $\Delta V$ are relatively small, such error sources can easily affect the estimation process of $\Delta V$. Although the estimated results of $\Delta V$ are inaccurate, their impacts on the joint estimation of SOC and capacity in the bias-compensated algorithm are minimal. For example, the estimated error for $\Delta V$ at the condition of ``Cell 1 at $\rm 25^\circ C$'' is -9.846mV when adding 10mV bias. In this condition, the slope of the OCV-SOC curve at 1\% SOC is 14.39V. When using DEKF to estimate the SOC, the estimated bias of the SOC caused by the inaccuracy of $\Delta V$ is just -0.07\% based on the calculation of \equref{E[e(k)]_f}, which can be ignored.

Although the estimated biases have errors, the proposed algorithm can track the bias changes quickly. When the measurement bias changes from 10mV to 30mV, the estimated bias in the third cycle also increases around 20mV compared to the previously estimated value in the second cycle. This is the real reason why the bias-compensated algorithm maintains its robustness under the scenario of bias mutation.

\section{Conclusions}\label{section 5}
The voltage measurement bias can reduce the accuracy of battery state estimation by generating an estimated bias for SOC. As a result of the analysis presented in this paper, this estimated bias becomes smaller when the slope of the OCV-SOC curve is larger. Following this insight, a bias-compensated algorithm has been proposed to reduce estimated errors related to SOC and SOH under the influence of voltage measurement bias. First, current signals at different frequencies and high-pass filters will be applied to estimate ohmic resistance, diffusion resistance, and time constant by the EKFs. After that, the estimated values of three parameters are used in the following estimation processes of battery states and voltage measurement bias. In the state estimation, DEKF estimates SOC and capacity only within the high-slope SOC region. Except for battery states, the voltage measurement bias is estimated in the SOC range with a low slope and then the estimated value will be compensated into the subsequent joint estimation of SOC and capacity to reduce the estimated errors of battery states further. The experimental results underscore the effectiveness of the proposed algorithm. Despite the large bias (30mV) in the voltage measurement and the bias mutation, the proposed algorithm is still robust and can accurately estimate battery states.
\bibliographystyle{IEEEtran}
\bibliography{main}
\end{document}